\documentclass[twocolumn,prb,showpacs,floats]{revtex4}
\usepackage[dvips]{graphicx}
\usepackage{dcolumn}
\usepackage{color}

\usepackage{multirow}

\usepackage{bm}

\newcommand{\CFOO}{Ca$_2$FeOsO$_6$}

\begin{document}

\title{Ferrimagnetism in the double perovskite Ca$_2$FeOsO$_6$: a density functional study}
\author{Hongbo Wang, Shasha Zhu, Xuedong Ou,}
\affiliation{Laboratory for Computational Physical Sciences (MOE), State Key Laboratory of Surface Physics, and Department of Physics, Fudan University, Shanghai 200433, China}
\author{Hua Wu}
\thanks{Corresponding author. wuh@fudan.edu.cn}
\affiliation{Laboratory for Computational Physical Sciences (MOE), State Key Laboratory of Surface Physics, and Department of Physics, Fudan University, Shanghai 200433, China}

\date{today}

\begin{abstract}

Using density functional calculations, we find that the newly synthesized Ca$_2$FeOsO$_6$ has the high-spin Fe$^{3+}$ ($3d^5$)-Os$^{5+}$ ($5d^3$) state. The octahedral Os$^{5+}$ ion has a large intrinsic exchange splitting, and its $t_{2g\uparrow}^3$ configuration makes the spin-orbit coupling ineffective. Moreover, there is a strong antiferromagnetic (AF) coupling between the neighboring Fe$^{3+}$ ($S$ = 5/2) and Os$^{5+}$ ($S$ = --3/2), but the AF couplings within both the fcc Fe$^{3+}$ and Os$^{5+}$ sublattices are one order of magnitude weaker. Therefore, a magnetic frustration is suppressed and a stable ferrimagnetic (FiM) ground state appears. This FiM order is due to the virtual hopping of the $t_{2g}$ electrons from Os$^{5+}$ ($t_{2g\downarrow}^3$) to Fe$^{3+}$ ($t_{2g\uparrow}^3e_{g\uparrow}^2$). However, if the experimental bended Fe$^{3+}$-O$^{2-}$-Os$^{5+}$ exchange path gets straight, the $e_g$ hopping from Fe$^{3+}$ ($t_{2g\uparrow}^3e_{g\uparrow}^2$) to Os$^{5+}$ ($t_{2g\uparrow}^3$) would be facilitated and then a ferromagnetic (FM) coupling would occur.

\end{abstract}


\maketitle
\section{Introduction}

Perovskites ABO$_3$ bear many functionalities such as colossal magnetoresistance and multiferroicity, which usually arise from their fascinating electronic and magnetic properties and the cross-coupling effects. Those properties, in a correlated electron system, often stem from an intriguing interplay among the charge, spin, and orbital degrees of freedom of the transition-metal (TM) cations and the lattice degree of freedom. Besides the widely studied $3d$ TM oxides, in recent years the $5d$ and $4d$ TM oxides also draw a lot of attention due to their significant spin-orbit coupling (SOC) effects~\cite{Kim08,Kim09}. Then crystal field, electron correlation, SOC, and band formation are all simultaneously present in such systems, and they would probably bring about novel electronic and magnetic properties~\cite{Jackeli,Wan,Mazin,Yin,Ou1,Ou2,Cao}.

The double perovskites A$_2$BB'O$_6$ (B=$3d$ TM, B'=$4d$ or $5d$ TM) could form a B-B' ordered atomic structure (interweaved fcc lattices) due to the quite different charge states and ionic sizes of the $3d$, and $4d$ or $5d$ TM cations~\cite{Anderson}. Sr$_2$FeMoO$_6$ and Sr$_2$FeReO$_6$ are two representative examples, and they are classified as a ferrimagnetic (FiM) half metal and display a giant tunneling magnetoresistance effect above room temperature~\cite{Tokura98,Tokura99}. Half metals bear fully spin-polarized carriers and are an important kind of spintronic materials and have promising technological applications. Therefore, double perovskites have been largely explored for their novel magnetic and electronic properties~\cite{Serrate,Alff,Meetei,Paul,Morrow,Yan,Feng}. For example, Sr$_2$CrOsO$_6$ shows a FiM order below $T_{\rm C}\sim$ 725 K, which seems to be the highest $T_{\rm C}$ record in the (double) perovskite oxides~\cite{Alff,Meetei}. Sr$_2$FeOsO$_6$ displays a lattice instability and competing spin structures~\cite{Paul}. In Sr$_2$CoOsO$_6$, the Co and Os sublattices exhibit different magnetic ground states and spin dynamics~\cite{Morrow,Yan}.

\begin{figure}[b]
\centering \includegraphics[width=7cm]{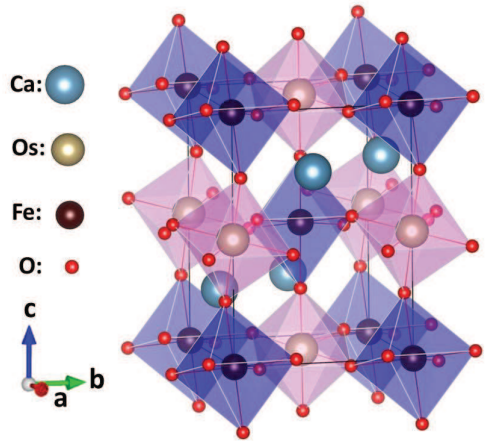}
\caption{
Double perovskite structure of {\CFOO}. Rotation and tilting of both the FeO$_6$ and OsO$_6$ octahedra cause bended
Fe-O-Os bonds, being 152$^{\circ}$ along the $c$ axis, and 152$^{\circ}$ and 154$^{\circ}$ in the $ab$ plane~\cite{Feng}.
}
\end{figure}

In this work, using density functional calculations, we study the electronic structure and magnetism of the double perovskite {\CFOO}, which was newly synthesized under a high- temperature and pressure~\cite{Feng}, see Fig. 1 for its crystal structure. It was reported to be a high-temperature FiM insulator with $T_{\rm C}\sim$ 320 K~\cite{Feng}.
As seen below, {\CFOO} has the high-spin Fe$^{3+}$-Os$^{5+}$ charge state. We plot in Fig. 2 a schematic level diagram and two possible exchange pathways, which lead to either an FM or an AF coupling between the Fe$^{3+}$ and Os$^{5+}$ ions. For a $d^5$-$d^3$ system, one may expect a FM coupling, taking into account a local Hund exchange and a $d^4$-$d^4$ charge fluctuation. In the present case, however, a FM superexchange would require a virtual hopping of the Fe$^{3+}$ $e_g$ electrons to the otherwise empty Os$^{5+}$ $e_g$ states, see Figs. 2(a) and 2(b). Then the excited Fe$^{4+}$-Os$^{4+}$ intermediate state is involved, but an involvement of the high-level Os $e_g$ crystal-field states could energetically disfavor this charge fluctuation process. On the other hand, when a more common Fe$^{2+}$-Os$^{6+}$ charge fluctuation state is considered, the down-spin Os$^{5+}$ $t_{2g}$ electrons would hop, forth and back, to the Fe$^{3+}$ $t_{2g}$ states, thus giving rise to an AF coupling between Fe$^{3+}$ and Os$^{5+}$, see Figs. 2(b) and 2(c). As a result, there could be a competition between the FM and AF couplings. Moreover, for both the magnetic Fe$^{3+}$ ($S$ = 5/2) and Os$^{5+}$ ($S$ = 3/2) sublattices, there is in principle an AF coupling in each fcc sublattice, and then spin frustration may also get involved. Therefore, the magnetic ground state of {\CFOO} and its spin-orbital physics are worth of a prompt study.

\begin{figure}[t]
\centering \includegraphics[width=6cm]{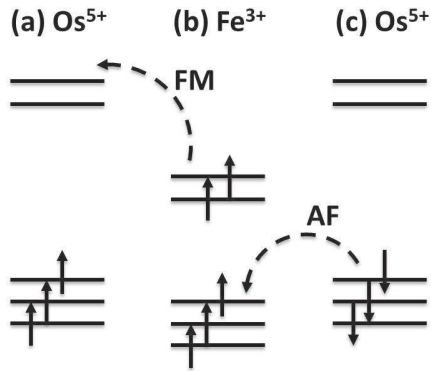}
\caption{
Schematic crystal field level diagrams of the high-spin Fe$^{3+}$ and Os$^{5+}$. While the $e_g$ electron hopping would lead to an FM coupling, the $t_{2g}$ electron hopping would result in an AF coupling.
}
\end{figure}

\section{Computational details}

We have used the full-potential augmented plane wave plus local orbital method coded in the Wien2k package~\cite{Blaha}. Experimental lattice parameters are used~\cite{Feng}: {\CFOO} is monoclinic ($\beta$ = 90.021$^{\circ}$, the space group $P2_1/n$) with the lattice constants $a$ = 5.393 \AA, $b$ = 5.508 \AA, and $c$ = 7.679 \AA. Owing to the rotation and tilting of the FeO$_6$ and OsO$_6$ octahedra, the Fe-O-Os bond angles range from 152 to 154$^{\circ}$, see Fig. 1. In the following calculations, we have used a local coordinate system with each axis being along the Fe-O (Os-O) bonds, in order to represent the Fe $3d$ and Os $5d$ orbital states in the local octahedral crystal field. Muffin-tin sphere radii are chosen to be 2.8, 2.2, and 1.4 bohr for Ca, Fe/Os, and O atoms. Cutoff energy of 16 Ryd is set for the interstitial plane wave expansion, and 6$\times$6$\times$4 k mesh for integration over the Brillouin zone. For the exchange-correlation functional, the generalized gradient approximation (GGA) is used; and GGA+$U$ calculations are also performed for including the static electron correlation~\cite{Anisimov}, with the Hubbard $U$ = 5 eV (2 eV) and Hund $J$ = 0.9 eV (0.4 eV) for Fe $3d$ (Os $5d$) states. The SOC is included by the second-variational method with scalar relativistic wave functions~\cite{Blaha}. As seen below, however, in {\CFOO} the Fe$^{3+}$ ($t_{2g}^3e_g^2$) and Os$^{5+}$ ($t_{2g}^3$) cations have a large exchange splitting and closed subshells. Therefore there is no orbital degree of freedom and the SOC is ineffective. Moreover, inclusion of the electron correlation does not affect the conclusions derived from the GGA calculations.

\section{Results and discussion}

\begin{figure}[t]
\centering \includegraphics[width=7cm]{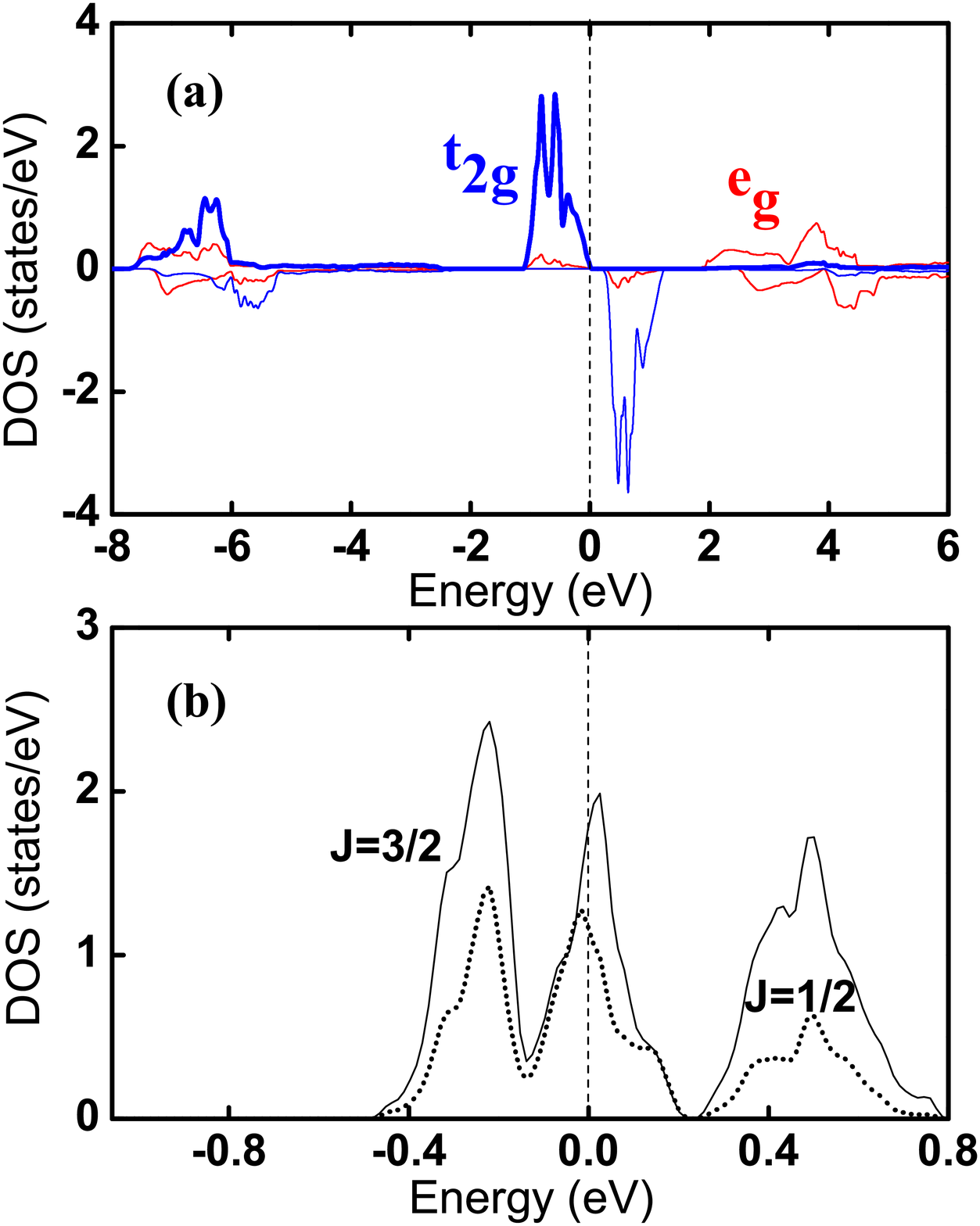}
\caption{(Color online)
(a) Os $5d$ density of states (DOS) calculated by spin-polarized GGA for the artificial Ca$_2$GaOsO$_6$. The Os$^{5+}$ ion has a fully spin-polarized $t_{2g\uparrow}^3$ configuration. (b) The $j$ = 3/2 and $j$ = 1/2 decomposed Os-$5d$ $t_{2g}$ DOS calculated by spin-restricted GGA+SOC. The strong mixing between the $j$ = 3/2 quartet and the $j$ = 1/2 doublet is due to the non-cubic crystal field and the band effect, see more in the main text.
}
\end{figure}

In the hybrid $3d$-5$d$ TM oxide {\CFOO}, the spins of the localized Fe $3d$ electrons would affect the delocalized Os $5d$ electrons. In order to see an intrinsic effect of the Os $5d$ electrons, we first carry out a spin-polarized GGA calculation for the artificial Ca$_2$GaOsO$_6$ (assuming the same structural parameters as {\CFOO}) to check a possible spin polarization of the Os ion, and a spin-restricted GGA+SOC calculation to see how strong the SOC is. We have replaced the magnetic Fe$^{3+}$ by nonmagnetic Ga$^{3+}$, both of which have a similar ionic size (0.645 \AA/Fe$^{3+}$ $vs$ 0.620 \AA/Ga$^{3+}$)~\cite{Shannon}. As seen in Fig. 3(a), the Os$^{5+}$ ion has a large $t_{2g}$-$e_g$ crystal field splitting of more than 3 eV, which would make only the $t_{2g}$ electrons active. The $t_{2g}$ states have an exchange splitting of 1.3 eV, and they are fully spin-polarized, with the formal $t_{2g\uparrow}^3$ configuration ($S$ = 3/2). Within its muffin-tin sphere, Os$^{5+}$ ion has a reduced spin moment of 1.81 $\mu_B$ due to the delocalized behavior of the $5d$ electrons and their strong covalency with the oxygen ligands (a moderate electron correlation slightly enhances the spin moment to 1.88 $\mu_B$ in the GGA+$U$ calculation for Ca$_2$GaOsO$_6$). For the spin-restricted GGA+SOC calculation, we have used the SOC basis set in a cubic crystal field, i.e., the $j$ = 3/2 quartet and $j$ = 1/2 doublet~\cite{Kim08}, to decompose the $t_{2g}$ DOS, see Fig. 3(b). However, we find a strong mixing between them, unlike the representative iridate Sr$_2$IrO$_4$ in which the $j$ = 3/2 and $j$ =1/2 states are well split by the SOC and they serve as a good basis set~\cite{Kim08,Ou1}. This difference can be attributed to the deviation from the cubic crystal field (due to the rotation and tilting of the OsO$_6$ and FeO$_6$ octahedra in {\CFOO}~\cite{Feng}), and to the inter-site hopping of the delocalized Os $5d$ electrons in the fcc sublattice. The non-cubic crystal field and the band effect smear out the otherwise SOC splitting of about 0.4 eV between the $j$ = 3/2 and $j$ = 1/2 states. Apparently, in {\CFOO} the Os$^{5+}$ ion has a large intrinsic exchange splitting and has a closed $t_{2g\uparrow}^3$ subshell. Owing to this and to the band formation, the SOC is insignificant, which can also be seen in the following GGA+$U$+SOC calculation.

\begin{table}[b]
\caption{Relative total energies (meV/fu) calculated by GGA and GGA+$U$. The G-AF (FiM) is the ground state.}
\begin{tabular}{l@{\hskip5mm}c@{\hskip5mm}c@{\hskip5mm}c@{\hskip5mm}c}
\hline\hline
  & FM & G-AF (FiM) & C-AF & A-AF \\ \hline
GGA & - & 0 & 90 & 273 \\
GGA+$U$ & 367 & 0 & 50 & 203  \\

\hline\hline
\end{tabular}
\end{table}

As the Os ion has an intrinsic spin polarization, now we study the exchange interaction between the Fe and Os ions in
{\CFOO}. Here we calculate, using GGA and GGA+$U$, the four magnetic states, namely, FM, G-AF (i.e., FiM with Fe$^{3+}$ being AF coupled to six neighboring Os$^{5+}$ and vice versa), C-AF (Fe$^{3+}$-Os$^{5+}$ AF coupling in the $ab$ plane and FM coupling along the $c$ axis), and A-AF (Fe$^{3+}$-Os$^{5+}$ FM coupling in the $ab$ plane and AF coupling along the $c$ axis). The total energy results are listed in Table I. Within GGA, the FM state is unstable and no converged solution can be achieved. From the total energy results of the three stable G-AF, C-AF, and A-AF solutions, we find the G-AF ground state with a local spin moment of 3.73 $\mu_B$/Fe$^{3+}$ and --1.43 $\mu_B$/Os$^{5+}$ (see Table II), which we can also name an FiM order. The G-AF ground state is insulating, with a band gap of 0.4 eV, see Fig. 4. We can see that in contrast to the large energy separation of more than 4 eV between the up-spin Fe$^{3+}$ and Os$^{5+}$ $e_{g}$ states, the down-spin Fe$^{3+}$ and Os$^{5+}$ $t_{2g}$ states are separated by only the small band gap of 0.4 eV, and their hybridization via the O $2p$ orbitals can push the occupied Os$^{5+}$ $t_{2g\downarrow}^3$ bands downwards and thus stabilizes the FiM ground state. In other words, the virtual hopping from Os$^{5+}$ $t_{2g\downarrow}^3$ to the empty Fe$^{3+}$ $t_{2g\downarrow}^0$ accounts for the FiM coupling and the decreasing spin moment of 1.43 $\mu_B$/Os$^{5+}$, compared with the above calculated 1.81 $\mu_B$/Os$^{5+}$ in the artificial Ca$_2$GaOsO$_6$ where no such hopping would occur. Taking into account the exchange couplings between the Fe$^{3+}$ and Os$^{5+}$ (nearest neighboring magnetic cations), and those of Fe$^{3+}$-Fe$^{3+}$ and of Os$^{5+}$-Os$^{5+}$ (next nearest neighbors) in each fcc sublattice, and neglecting the exchange anisotropy, we can write the energy associated with different magnetic order to be \\

FM: 6$J_{\rm Fe-Os}$ + 6$J_{\rm Fe-Fe}$ + 6$J_{\rm Os-Os}$ per formula unit, \\

G-AF: --6$J_{\rm Fe-Os}$ + 6$J_{\rm Fe-Fe}$ + 6$J_{\rm Os-Os}$, \\

C-AF: --2$J_{\rm Fe-Os}$ -- 2$J_{\rm Fe-Fe}$ -- 2$J_{\rm Os-Os}$, \\

and A-AF: 2$J_{\rm Fe-Os}$ -- 2$J_{\rm Fe-Fe}$ -- 2$J_{\rm Os-Os}$.\\

\noindent{Here} $J_{\rm Fe-Os}$, $J_{\rm Fe-Fe}$, and $J_{\rm Os-Os}$
stand for the exchange energy for each Fe-Os, Fe-Fe, and Os-Os pair, respectively. Then the energy difference between the C-AF and A-AF states, 273 -- 90 = 183 meV/fu (equal to 4$J_{\rm Fe-Os}$), allows us to estimate the AF exchange energy per Fe-Os pair ($J_{\rm Fe-Os}$) to be about 46 meV.

\begin{figure}[t]
\centering \includegraphics[width=7cm]{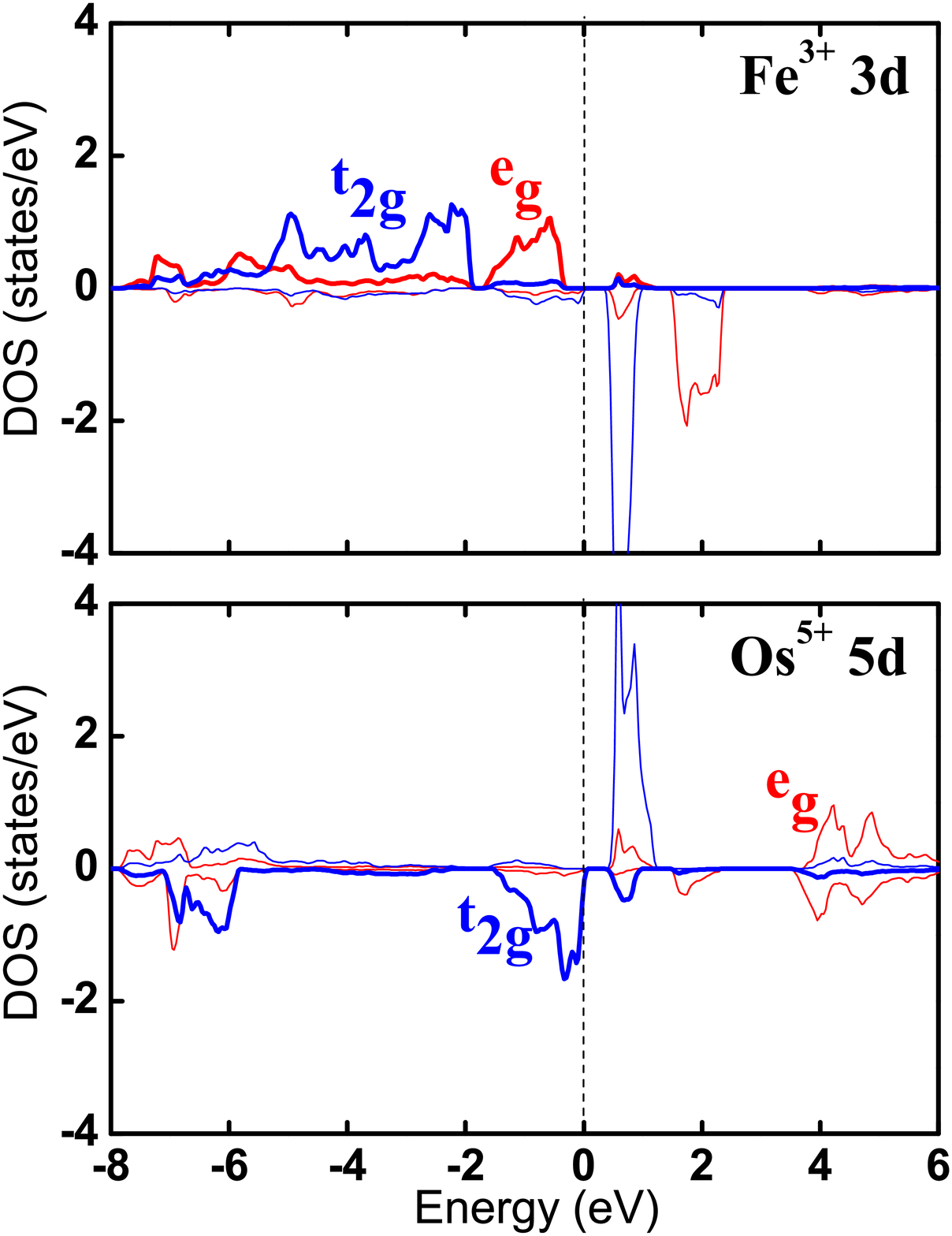}
\caption{(Color online)
Fe$^{3+}$ $3d$ and Os$^{5+}$ $5d$ DOS calculated by spin-polarized GGA for the FiM ground state. Positive (negative) DOS values stand for the up (down) spin channel. Fermi level is set at zero energy. The down-spin Fe$^{3+}$-Os$^{5+}$ $t_{2g}$ states are separated by only a small gap of 0.4 eV, and their hybridization stabilizes the FiM ground state.
}
\end{figure}

The above GGA results show that {\CFOO} is a narrow band insulator. A strong (moderate) electron correlation could well be present for the Fe $3d$ (Os $5d$) states. As such, we also carry out GGA+$U$ calculations. We find in Table I that the G-AF (FiM) state remains to be the ground state, with now increasing spin moment of 4.13 $\mu_B$/Fe$^{3+}$ and --1.72 $\mu_B$/Os$^{5+}$ (see Table II). The energy difference between the C- and G-AF and that between the A- and G-AF are reduced due to the correlation driven electron localization and decreasing exchange interactions. Note that owing to the strong on-site and orbitally polarized Coulomb potential exerted by the +$U$ method, the FM state can now be stabilized. In this FM state, see Figs. 2(a) and 2(b), the Os$^{5+}$ ($t_{2g\uparrow}^3$)-Fe$^{3+}$ electron hopping is forbidden, but the Fe$^{3+}$ ($e_g^2$)-Os$^{5+}$ electron hopping is facilitated, both of which contribute to the increasing Os$^{5+}$ spin moment of 1.96 $\mu_B$, compared with 1.72 $\mu_B$ in the G-AF ground state of {\CFOO} and with 1.81-1.88 $\mu_B$ in the artificial Ca$_2$GaOsO$_6$. However, this metastable FM state lies at a much higher energy than the G-AF ground state by 367 meV/fu (equal to 12$J_{\rm Fe-Os}$). This allows us to estimate the decreasing Fe$^{3+}$-Os$^{5+}$ AF exchange energy ($J_{\rm Fe-Os}$) to be about 31 meV (compared with 46 meV by GGA). Note that the AF $J_{\rm Fe-Os}$ can also be estimated, from the energy difference between the C- and A-AF states (203 --50 = 153 meV/fu, equal to 4$J_{\rm Fe-Os}$), to be about 38 meV. The difference (31 $vs$ 38 meV) is reasonable, as the anisotropic exchange constants in this distorted {\CFOO} are modeled by a single parameter for each Fe-Os, Fe-Fe and Os-Os pair. As seen in Fig. 5, the insulating gap of the G-AF (FiM) ground state is increased up to 1.1 eV due to the electron correlation effect. This G-AF (FiM) insulating ground-state solution agrees well with the most recent experimental finding~\cite{Feng}.

\begin{table}[t]
\caption{Spin magnetic moments (in unit of $\mu_B$) of Fe$^{3+}$, Os$^{5+}$, and apical and planar O$^{2-}$ ions in the FiM ground state of {\CFOO}, and the insulating band gap (in unit of eV). Os$^{5+}$ has also a small orbital moment of 0.13 $\mu_B$ by GGA+$U$+SOC.}
\begin{tabular}{l@{\hskip3mm}c@{\hskip3mm}c@{\hskip3mm}c@{\hskip3mm}c@{\hskip3mm}c}
\hline\hline
  & Fe & Os & $c$-O & $ab$-O & Gap\\ \hline
GGA & 3.73 & --1.43 & 0.01 & --0.02 & 0.4 \\
GGA+$U$ & 4.13 & --1.72 & 0 & --0.03 & 1.1  \\
GGA+$U$+SOC & 4.13 & --1.64 (0.13) & 0 & --0.03 & 0.8  \\

\hline\hline
\end{tabular}
\end{table}

\begin{figure}[t]
\centering \includegraphics[width=7cm]{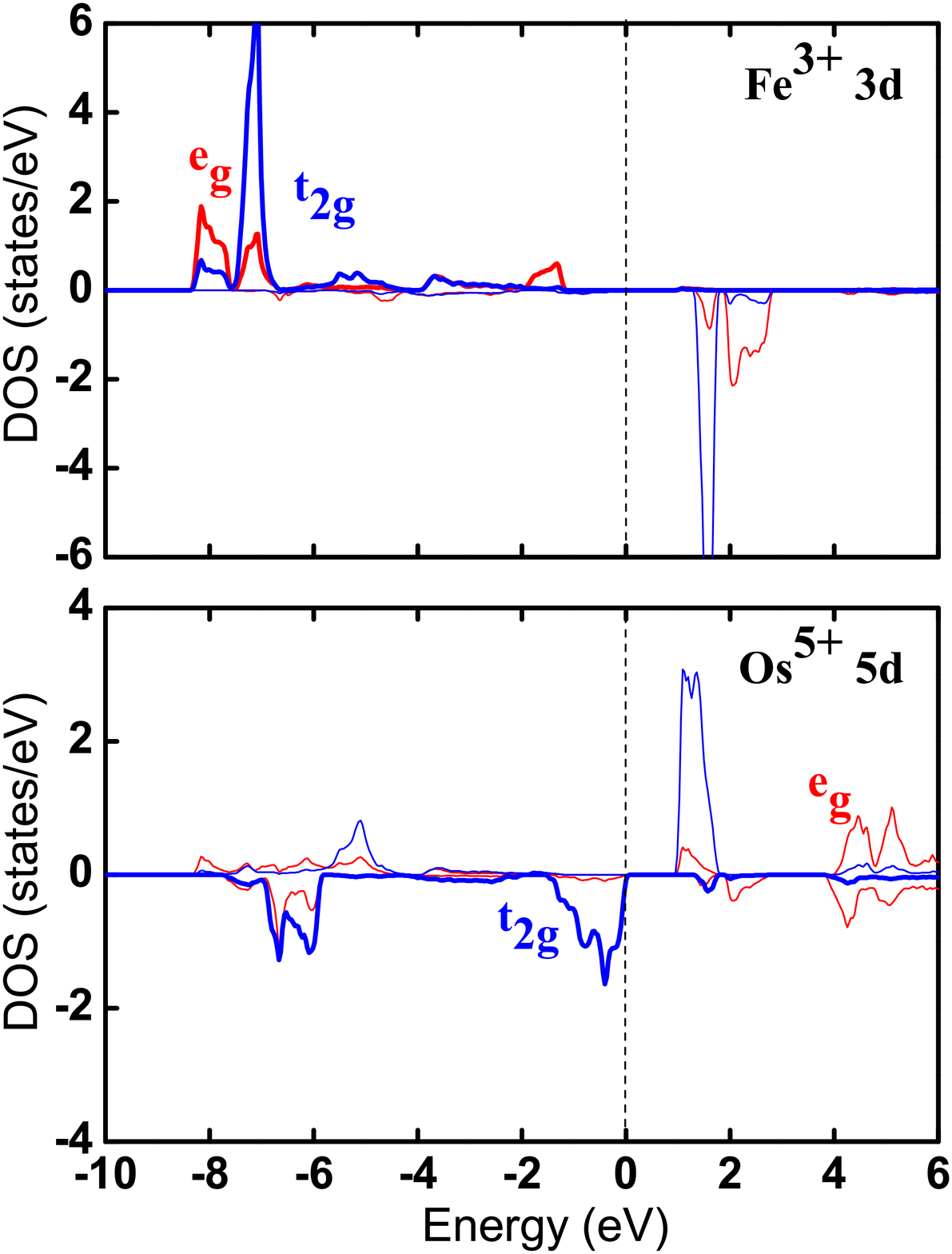}
\caption{(Color online)
Fe$^{3+}$ $3d$ and Os$^{5+}$ $5d$ DOS calculated by GGA+$U$ for the FiM ground state. The insulating gap is increased to 1.1 eV.
}
\end{figure}

Now we probe the possible AF coupling within both the fcc Fe$^{3+}$ and Os$^{5+}$ sublattices, whose values however cannot be extracted from the energy differences of those magnetic states listed in Table I. We calculate the FM state and the layered AF state (FM $ab$ planes in an AF stacking along the $c$ axis) for both the artificial Ca$_2$GaOsO$_6$ and La$_2$FeGaO$_6$ (both assuming the same structural parameters as {\CFOO}, and moreover, 0.645 \AA/Fe$^{3+}$, 0.575 \AA/Os$^{5+}$, and 0.620 \AA/Ga$^{3+}$ all in a similar size~\cite{Shannon}). Then $J_{\rm Os-Os}$ and $J_{\rm Fe-Fe}$ can be separately estimated. GGA+$U$ calculations ($U$ = 2 eV and Hund $J$ = 0.4 eV for Os $5d$ states) for Ca$_2$GaOsO$_6$ show that the layered AF state is more stable than the FM state by 27 meV/fu (equal to 8$J_{\rm Os-Os}$), which allows us to estimate the AF $J_{\rm Os-Os}$ to be about 3.5 meV. Corresponding GGA+$U$ calculations ($U$ = 5 eV and Hund $J$ = 0.9 eV for Fe $3d$ states) for La$_2$FeGaO$_6$ show that the layered AF state is more stable than the FM state by 20 meV/fu (equal to 8$J_{\rm Fe-Fe}$). Then the AF $J_{\rm Fe-Fe}$ is estimated to be 2.5 meV. Compared with the delocalized Os$^{5+}$ $5d$ orbitals (formal $S$ = 3/2) and probably stronger Os-Os magnetic coupling, the localized Fe$^{3+}$ $3d$ orbitals ($S$ = 5/2) have a smaller exchange constant, but the larger spin = 5/2 compensates for the weaker exchange, giving rise to the comparable AF coupling energy for each Os-Os pair and the Fe-Fe one (3.5 meV $vs$ 2.5 meV). 

With $J_{\rm Fe-Os}$ = 31 meV, $J_{\rm Os-Os}$ = 3.5 meV, and $J_{\rm Fe-Fe}$ = 2.5 meV, in {\CFOO} the C-AF state can also be estimated to be higher in energy than the G-AF ground state by 4$J_{\rm Fe-Os}$ -- 8$J_{\rm Os-Os}$ -- 8$J_{\rm Fe-Fe}$ = 76 meV/fu, and the A-AF state is less stable than the G-AF by 8$J_{\rm Fe-Os}$ -- 8$J_{\rm Os-Os}$ -- 8$J_{\rm Fe-Fe}$ = 200 meV/fu. Besides 367 meV/fu energy difference between the FM and G-AF states (equal to 12$J_{\rm Fe-Os}$), the GGA+$U$ calculated energy difference of 50 meV/fu between the C-AF and G-AF (see Table I), and that of 203 meV/fu between the A-AF and G-AF are both well comparable or close to the two estimated values of 76 and 200 meV/fu. This also justifies our choice of the artificial Ca$_2$GaOsO$_6$ and La$_2$FeGaO$_6$ to estimate the AF $J_{\rm Os-Os}$ and $J_{\rm Fe-Fe}$. As the nearest neighboring Fe$^{3+}$-Os$^{5+}$ AF coupling energy (more than 30 meV per pair) is one order of magnitude larger than the next nearest neighboring Os-Os and Fe-Fe AF coupling energy (both with only few meV), a possible magnetic frustration in both the fcc Fe$^{3+}$ and Os$^{5+}$ sublattices is actually suppressed, and then the collinear FiM insulating state turns out to be the ground state. Using a simple mean-field model, $T_{\rm C}$ could be estimated to be $ZJ_{\rm Fe-Os}$/(3$k_{\rm B}$) = 2$J_{\rm Fe-Os}$/$k_{\rm B}$ $\sim$ 600 K, which would decrease upon a spin fluctuation and the possible spin frustration. This estimated $T_{\rm C}$ and the calculated band gap of 1.1 eV for {\CFOO} well account for the experimental above-room-temperature $T_{\rm C}$ = 320 K and the insulating gap of 1.2 eV~\cite{Feng}.

Finally, we note that in the present case, the SOC is insignificant, as our GGA+$U$+SOC calculation for the FiM ground state shows that the Os$^{5+}$ has a spin moment of --1.64 $\mu_B$ (compared with --1.73 $\mu_B$ by GGA+$U$) but a small orbital moment of 0.13 $\mu_B$. Moreover, we explore a possibility to get a FM Fe$^{3+}$-Os$^{5+}$ coupling, which might be expected for a $d^5$-$d^3$ pair due to the $e_g$ hopping mechanism. In the test calculations, we stretch the $c$-axis Fe-O-Os bended bonds (152$^{\circ}$) into straight ones by slightly enlarging the $c$-axis lattice constant (from the experimental 7.68 \AA~\cite{Feng} to the artificial 7.92 \AA) and moving apical oxygens into the straight bonds (for fixing the experimental Fe-O-Os bond distances). As the $c$-axis straight bonds would facilitate the $e_g$ hopping, a $c$-axis FM coupling could be promoted and then the C-AF state could become most stable. Indeed, our calculations show that with such structural changes, the C-AF becomes the ground state and it is more stable than the G-AF state by 31 meV/fu in GGA and by 35 meV/fu in GGA+$U$. Note, however, that a large elastic energy cost of about 1.2 eV/fu, associated with the structural changes, would actually prevent such phase occurring. Then G-AF (FiM) is indeed the magnetic ground state for {\CFOO}.

\section{Conclusion}

In summary, we find, using density functional calculations, that the newly synthesized double perovskite {\CFOO} is an FiM insulator. It has the high-spin Fe$^{3+}$-Os$^{5+}$ charge state. The Os$^{5+}$ ion has a large exchange splitting and has a closed $t_{2g\uparrow}^3$ subshell, and therefore the spin-orbit coupling is insignificant. The nearest neighboring Fe$^{3+}$-Os$^{5+}$ AF coupling energy is more than 30 meV per pair, but the next nearest neighboring AF coupling energy per Os$^{5+}$-Os$^{5+}$ pair and that per Fe$^{3+}$-Fe$^{3+}$ pair are both only few meV. Thus, a possible magnetic frustration is suppressed in both the fcc Os$^{5+}$ and Fe$^{3+}$ sublattices, and {\CFOO} displays a high-$T_{\rm C}$ FiM order. This work well accounts for the recent experiment~\cite{Feng}.\\

{\bf Acknowledgment.}
This work was supported by the NSF of China (Grant No. 11274070),
PuJiang Program of Shanghai (Grant No. 12PJ1401000), and
ShuGuang Program of Shanghai (Grant No. 12SG06).


\begin{thebibliography}{22}

\bibitem{Kim08}
B. J. Kim, Hosub Jin, S. J. Moon, J.-Y. Kim, B.-G. Park, C. S. Leem,
J. Yu, T. W. Noh, C. Kim, S.-J. Oh, J.-H. Park, V. Durairaj, G. Cao,
and E. Rotenberg, Phys. Rev. Lett. {\bf 101}, 076402 (2008).

\bibitem{Kim09}
B. J. Kim, H. Ohsumi, T. Komesu, S. Sakai, T. Morita, H. Takagi, and T. Arima,
Science {\bf 323}, 1329 (2009).

\bibitem{Jackeli}
G. Jackeli and G. Khaliullin, Phys. Rev. Lett. {\bf 102}, 017205 (2009).

\bibitem{Wan}
X. G. Wan, A. M. Turner, A. Vishwanath, and S. Y. Savrasov,
Phys. Rev. B {\bf 83}, 205101 (2011).

\bibitem{Mazin}
I. I. Mazin, Harald O. Jeschke, K. Foyevtsova, Roser Valent\'i,
and D. I. Khomskii, Phys. Rev. Lett. {\bf 109}, 197201 (2012).

\bibitem{Yin}
W. G. Yin, X. Liu, A. M. Tsvelik, M. P. M. Dean, M. H. Upton, J. Kim, D. Casa,
A. Said, T. Gog, T. F. Qi, G. Cao, and J. P. Hill,
Phys. Rev. Lett. {\bf 111}, 057202 (2013).

\bibitem{Ou1}
X. Ou and H. Wu, Phys. Rev. B {\bf 89}, 035138 (2014).

\bibitem{Ou2}
X. Ou and H. Wu, Sci. Rep. {\bf 4}, 4609 (2014).

\bibitem{Cao}
G. Cao, T. F. Qi, L. Li, J. Terzic, S. J. Yuan, L. E. DeLong, G. Murthy, and R. K. Kaul,
Phys. Rev. Lett. {\bf 112}, 056402 (2014).

\bibitem{Anderson}
M.T. Anderson, K.B. Greenwood, G.A. Taylor, and K.R. Poeppelmeier,
Prog. Solid State Chem. {\bf 22}, 197 (1993).

\bibitem{Tokura98}
K.-I. Kobayashi, T. Kimura, H. Sawada, K. Terakura, and Y. Tokura,
Nature (London) {\bf 395}, 677 (1998).

\bibitem{Tokura99}
K.-I. Kobayashi, T. Kimura, Y. Tomioka, H. Sawada, K. Terakura, and Y. Tokura,
Phys. Rev. B {\bf 59}, 11159 (1999).

\bibitem{Serrate}
D. Serrate, J. M. De Teresa, and M. R. Ibarra,
J. Phys. Condens. Matter {\bf 19}, 023201 (2007).

\bibitem{Alff}
Y. Krockenberger, K. Mogare, M. Reehuis, M. Tovar, M. Jansen, G. Vaitheeswaran, V. Kanchana,
F. Bultmark, A. Delin, F. Wilhelm, A. Rogalev, A. Winkler, and L. Alff,
Phys. Rev. B {\bf 75}, 020404(R) (2007).

\bibitem{Meetei}
O. N. Meetei, O. Erten, M. Randeria, N. Trivedi, and P. Woodward,
Phys. Rev. Lett. {\bf 110}, 087203 (2013).

\bibitem{Paul}
A. K. Paul, M. Reehuis, V. Ksenofontov, B. Yan, A. Hoser, D. M. T\"{o}bbens, P. M. Abdala,
P. Adler, M. Jansen, and C. Felser,
Phys. Rev. Lett. {\bf 111}, 167205 (2013).

\bibitem{Morrow}
R. Morrow, R. Mishra, O. D. Restrepo, M. R. Ball, W. Windl, S. Wurmehl, U. Stockert,
B. B\"{u}chner, and P. M. Woodward, J. Am. Chem. Soc. {\bf 135}, 18824 (2013).

\bibitem{Yan}
B. Yan, A. K. Paul, S. Kanungo, M. Reehuis, A. Hoser, D. M. T\"{o}bbens, W. Schnelle, R. C. Williams,
T. Lancaster, F. Xiao, J. S. M\"{o}ller, S. J. Blundell, W. Hayes, C. Felser, and M. Jansen,
Phys. Rev. Lett. {\bf 112}, 147202 (2014).

\bibitem{Feng}
H. L. Feng, M. Arai, Y. Matsushita, Y. Tsujimoto, Y. Guo, C. I. Sathish, X. Wang,
Y. H. Yuan, M. Tanaka, and K. Yamaura, J. Am. Chem. Soc. {\bf 136}, 3326 (2014).

\bibitem{Blaha} P. Blaha, K. Schwarz, G. Madsen, D. Kvasnicka, and J. Luitz,
{\bf WIEN2k}, 2001. ISBN 3-9501031-1-2.

\bibitem{Anisimov} V. I. Anisimov, I. V. Solovyev, M. A. Korotin,
M. T. Czy\.zyk, and G. A. Sawatzky, Phys. Rev. B {\bf 48}, 16929 (1993).

\bibitem{Shannon}
R. D. Shannon, Acta Crystallogr. Sect. A {\bf 32}, 751 (1976).



\end{thebibliography}
\end{document}